\documentclass[11pt]{article}

\usepackage[utf8]{inputenc}
\usepackage{amsmath}
\usepackage{graphicx}
\usepackage[hidelinks]{hyperref}
\usepackage{cite}
\usepackage{booktabs}
\usepackage[a4paper, total={6.5in, 9in}]{geometry} 
\usepackage{subcaption}

\title{\textbf{PIR-RAG: A System for Private Information Retrieval in Retrieval-Augmented Generation}}

\author{
  Baiqiang Wang\textsuperscript{1}, Qian Lou\textsuperscript{2}, Mengxin Zheng\textsuperscript{2}, Dongfang Zhao\textsuperscript{1} \\
  \textsuperscript{1}University of Washington \\
  \texttt{\{wbq@uw.edu, dzhao@cs.washington.edu\}} \\
  \and
  \textsuperscript{2}University of Central Florida \\
  \texttt{\{qian.lou, mengxin.zheng\}@ucf.edu}
}

\date{}

\begin{document}

\maketitle

\begin{abstract}
Retrieval-Augmented Generation (RAG) has become a foundational component of modern AI systems, yet it introduces significant privacy risks by exposing user queries to service providers. To address this, we introduce PIR-RAG, a practical system for privacy-preserving RAG. PIR-RAG employs a novel architecture that uses coarse-grained semantic clustering to prune the search space, combined with a fast, lattice-based Private Information Retrieval (PIR) protocol. This design allows for the efficient retrieval of entire document clusters, uniquely optimizing for the end-to-end RAG workflow where full document content is required. Our comprehensive evaluation against strong baseline architectures, including graph-based PIR and Tiptoe-style private scoring, demonstrates PIR-RAG's scalability and its superior performance in terms of "RAG-Ready Latency"—the true end-to-end time required to securely fetch content for an LLM. Our work establishes PIR-RAG as a viable and highly efficient solution for privacy in large-scale AI systems.
\end{abstract}

\noindent\textbf{Keywords:} Privacy-Preserving Machine Learning, Private Information Retrieval, Retrieval-Augmented Generation, Homomorphic Encryption.

\section{Introduction}
The rapid proliferation of Large Language Models (LLMs) is fundamentally reshaping how users interact with information. A key catalyst in this transformation is the Retrieval-Augmented Generation (RAG) paradigm \cite{lewis2020}, which grounds LLMs in vast external knowledge bases, enabling them to provide factually accurate, up-to-date, and contextually relevant responses. From enterprise knowledge management systems to personal AI assistants, RAG has become the de facto standard for building powerful, knowledgeable AI applications.

However, this powerful architecture introduces a critical and often overlooked privacy dilemma. The retrieval step, which is the cornerstone of RAG, requires users to send their queries—or high-fidelity vector embeddings of them—to a centralized server that hosts the document corpus. These queries can be deeply personal, commercially sensitive, or legally privileged, ranging from health inquiries and financial planning questions to proprietary research and development topics. Exposing this information to the service provider creates a significant privacy risk, undermining user trust and creating barriers to adoption in sensitive domains.

The natural cryptographic solution to this problem is Private Information Retrieval (PIR), a protocol that allows a client to fetch a data item from a server without revealing which item was chosen. Despite recent and significant performance improvements driven by lattice-based cryptography \cite{henzinger2023one}, a naive application of PIR to RAG systems remains computationally infeasible. A single PIR query over a database of millions of documents is still too slow and resource-intensive for an interactive user experience.

This performance bottleneck has spurred the development of specialized private search architectures designed to make PIR practical. These systems typically first prune the vast search space before invoking a cryptographic protocol. Graph-based architectures, such as PACMANN \cite{pacmann2024}, construct intricate similarity graphs to enable highly accurate, targeted retrievals, but often at the cost of complex, heavyweight client-side preprocessing. A more lightweight alternative is the cluster-based paradigm, used by systems like Tiptoe \cite{tiptoe2023}, which partitions the database and uses fast homomorphic encryption to return only a ranked list of identifiers. While efficient, this approach is fundamentally misaligned with the end goal of RAG, which requires the full content of the retrieved documents, not just their IDs. Retrieving this content would necessitate a series of subsequent, expensive PIR queries, nullifying the initial performance gains.

In this work, we argue that a truly practical solution for private RAG requires a new architecture designed from the ground up for its specific workflow. We introduce PIR-RAG, a novel system that addresses this need. Our core insight is that the architectural choice of retrieving an entire pre-defined document cluster in a single, batched PIR operation is uniquely suited for RAG. While seemingly coarse-grained, this "cluster-and-fetch" design elegantly amortizes the cost of content fetching, which is an unavoidable and expensive final step in any other private retrieval architecture. By merging the retrieval and fetching phases, PIR-RAG transforms what would be K+1 slow operations into a single, highly efficient one.

We summarize our contributions as follows:
\begin{itemize}
    \item We propose a new architecture for private RAG that combines coarse-grained semantic clustering with a fast, lattice-based PIR protocol, specifically optimizing for the end-to-end content retrieval task.
    \item We present an efficient system implementation featuring a novel chunk-transposed data structure, which transforms the private retrieval into a single, high-performance homomorphic matrix-vector product.
    \item We conduct a comprehensive comparative evaluation of PIR-RAG against our own implementations of strong graph-based and Tiptoe-style baselines, providing a clear, data-driven analysis of the architectural trade-offs in terms of scalability, performance, and search quality. Our results establish PIR-RAG as a highly practical and efficient solution for building the next generation of privacy-preserving AI systems.
\end{itemize}

Our main contributions are:
\begin{itemize}
    \item \textbf{A New Architecture for Private RAG:} We propose a "cluster-and-fetch" architecture that uses coarse-grained semantic indexing to prune the search space, making PIR computationally practical.
    \item \textbf{An Efficient System Implementation:} We build our system using a fast, lattice-based homomorphic encryption scheme and a novel chunk-transposed data structure that transforms retrieval into a single, highly efficient matrix-vector product.
    \item \textbf{A Comprehensive Comparative Evaluation:} We benchmark PIR-RAG against our own implementations of two powerful baseline architectures: a graph-based PIR system (Graph-PIR) and a Tiptoe-style private scoring system, providing a clear analysis of the trade-offs in the design space.
\end{itemize}

Our results demonstrate that PIR-RAG's architecture provides a compelling balance of scalability, performance, and privacy, proving particularly effective for the end-to-end RAG task.

\section{Related Work}
Our work lies at the intersection of Retrieval-Augmented Generation (RAG), Private Information Retrieval (PIR), and the system architectures that make their combination practical. We situate PIR-RAG by reviewing advances in each of these areas.

\subsection{Retrieval-Augmented Generation and its Privacy Gap}
Retrieval-Augmented Generation (RAG) \cite{lewis2020} has emerged as a dominant paradigm for enhancing Large Language Models (LLMs) with timely and factual knowledge from external sources. The standard RAG workflow involves embedding a user's query, searching a vector database for relevant document chunks, and providing these chunks as context to an LLM. While highly effective at improving factual grounding and reducing hallucinations, this process introduces a critical privacy vulnerability: the retrieval server is exposed to the user's queries (or their embeddings), which can reveal sensitive personal, commercial, or proprietary information. Our work directly addresses this privacy gap by designing a system where the retrieval step is cryptographically protected.

\subsection{Advances in Private Information Retrieval}
Private Information Retrieval (PIR) is a cryptographic protocol, first introduced by Chor et al. \cite{chor1998}, that allows a client to retrieve an item from a server's database without the server learning which item was retrieved. Historically, PIR schemes were considered too computationally expensive for practical use. However, recent breakthroughs have drastically improved their performance, making them viable for real-world systems.

Modern PIR schemes are often built on homomorphic encryption. The field has increasingly shifted towards schemes built on lattice-based cryptography, specifically the Learning With Errors (LWE) problem \cite{regev2009}. These LWE-based linearly homomorphic encryption (LHE) schemes, such as those used in SimplePIR \cite{henzinger2023one} and YPIR \cite{menon2024}, offer computational speeds that are orders of magnitude faster than predecessors. The core computation, often a matrix-vector product, becomes almost as fast as its plaintext equivalent, shifting the primary performance bottleneck from computation to communication. Our PIR-RAG system leverages a fast, LWE-based PIR protocol to ensure high server-side performance.

\subsection{Architectures for Private Search}
Applying a PIR primitive naively to a large database remains inefficient. The key research challenge, therefore, lies in designing system architectures that intelligently structure data to make PIR practical. Prior work on private document retrieval, such as Coeus \cite{ahmad2021}, focused on oblivious ranking using TF-IDF, but modern systems increasingly rely on semantic search over vector embeddings. We categorize recent semantic search architectures into two relevant paradigms.

\textbf{Cluster-based Architectures.} This approach, which both PIR-RAG and Tiptoe \cite{tiptoe2023} belong to, involves partitioning the database into semantic clusters. The client first identifies a target cluster locally using public metadata (e.g., centroids) and then uses a PIR protocol to interact only with that cluster's data. This dramatically prunes the search space. However, the architectural goals of Tiptoe and PIR-RAG diverge significantly. Tiptoe is designed for private web search, where the goal is to return a ranked list of identifiers or scores. It uses its LHE scheme to privately compute similarity scores on the server, returning only encrypted scores. This makes it incomplete for RAG, which requires the full document content. PIR-RAG, in contrast, is architecturally designed for the RAG workflow. It uses the PIR protocol to retrieve the \textit{entire content} of the target cluster in a single operation, thereby merging the "retrieval" and "fetching" steps.

\textbf{Graph-based Architectures.} This paradigm, exemplified by systems like PACMANN \cite{pacmann2024} and Compass \cite{compass2024}, prioritizes search accuracy by building a k-nearest neighbor (k-NN) similarity graph, often using the HNSW algorithm \cite{malkov2018}. Retrieval is performed via a private, multi-step traversal of this graph. These systems can achieve higher search quality because their search is not confined to a single cluster. However, they come with their own distinct trade-offs. PACMANN, for instance, achieves extremely low online query latency but does so by offloading a massive preprocessing burden to the client, which must stream the entire index. Compass uses Oblivious RAM (ORAM) \cite{stefanov2013} to achieve even stronger privacy by hiding access patterns, but this comes at a significant performance cost.

PIR-RAG carves out a unique position in this design space. It avoids the heavy client-side burden of systems like PACMANN and, unlike scoring-based systems like Tiptoe, is architecturally tailored for the end-to-end RAG task. By efficiently delivering full document content in a single private operation, it offers a practical and balanced solution specifically for privacy-preserving RAG.

\section{System Design: PIR-RAG}
\begin{figure*}[ht!]
  \centering
  \includegraphics[width=0.9\textwidth]{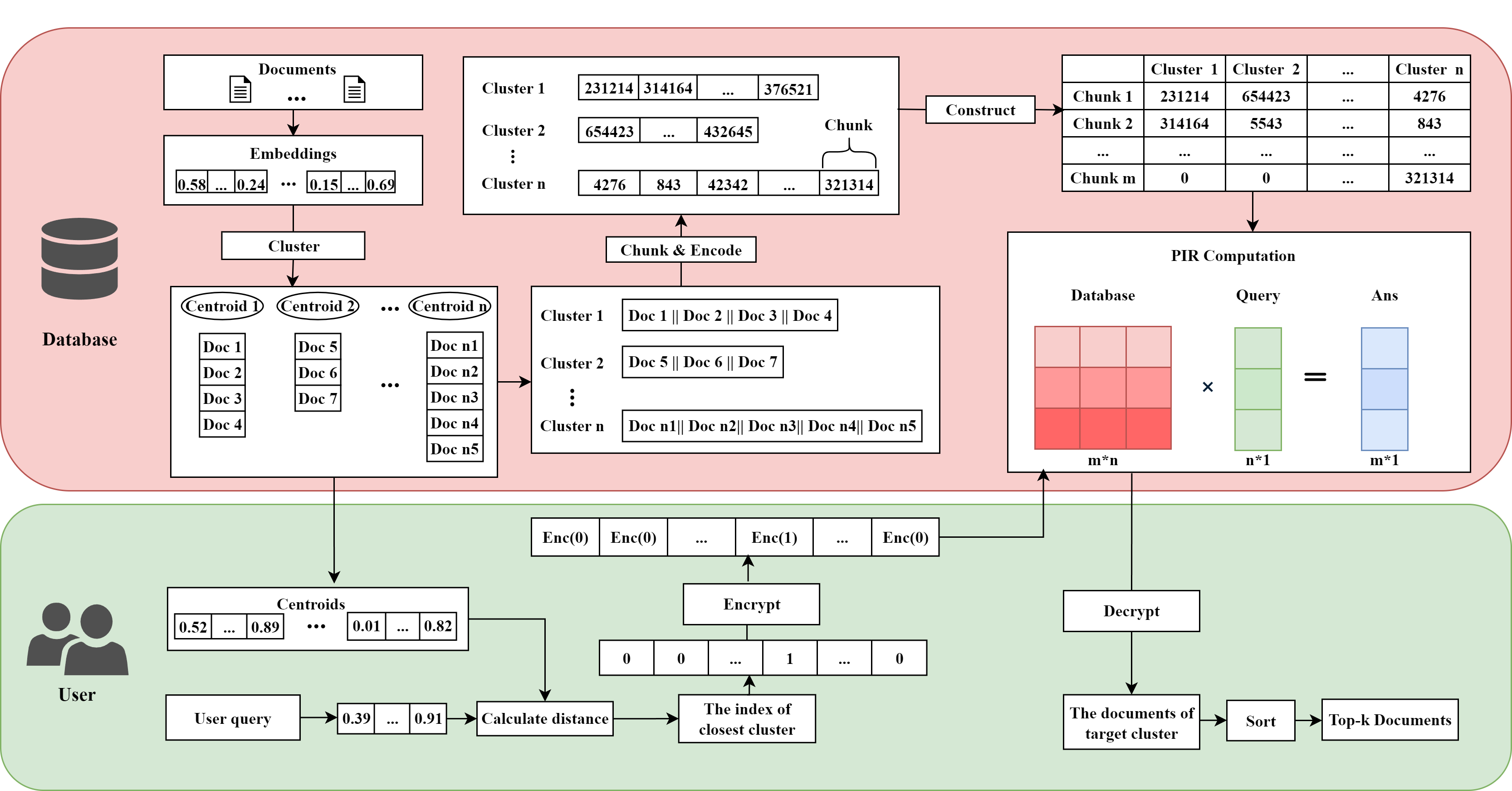} 
  \caption{The PIR-RAG system architecture. (1) Offline Setup: The server partitions the document corpus into semantic clusters. (2) Online Query: The client identifies the closest cluster locally and uses a fast LWE-based PIR scheme to privately download the entire cluster for local re-ranking.}
  \label{fig:system}
\end{figure*}

We designed PIR-RAG to enable private document retrieval in two primary phases: an offline setup phase and an online query phase. The overall architecture is shown in Figure~\ref{fig:system}.

\subsection{Threat Model}
We operate under the honest-but-curious server model. The server will correctly follow the protocol but may attempt to analyze all data it observes (i.e., the user's encrypted queries) to infer information about the user's interests. The goal of PIR-RAG is to ensure the server learns nothing about the specific document cluster the user is interested in.

\subsection{Offline Phase: Corpus Preprocessing}
Before any queries can be served, the server preprocesses the document corpus in a one-time setup consisting of two stages.

\textbf{Embedding and Clustering.} Each document in the corpus is converted into a numerical vector embedding using a standard model (e.g., \texttt{bge-base-en-v1.5}). The server then applies a clustering algorithm, such as K-means, to group these embeddings into $n$ semantic clusters. The centroids of these clusters are made public for client-side use.

\textbf{Chunk-Transposed Database Construction.} For each cluster, the documents are concatenated and partitioned into fixed-size chunks. This collection of chunks from all clusters is then organized into a chunk-transposed matrix of size $m \times n$, where $m$ is the number of chunks per cluster and $n$ is the number of clusters. This structure is the key to our performance, as it allows the retrieval of an entire cluster's documents to be formulated as a single matrix-vector multiplication.

\subsection{Online Phase: Private Retrieval}
When a user wants to issue a query, the online private retrieval process unfolds in several key stages.

\textbf{Client-Side Query Formulation.} The user's client software first computes the embedding for the query. It then compares this embedding to the public centroids of the $n$ clusters and identifies the index $i$ of the closest cluster. Following this, the client constructs an $n$-dimensional one-hot vector with a $1$ at the identified index $i$ and encrypts it using a fast, lattice-based scheme. This encrypted query is the only information sent to the server.

\textbf{Server-Side Private Computation.} The server receives the encrypted query vector. It then performs a highly efficient homomorphic matrix-vector multiplication between its public $m \times n$ database and the user's encrypted $n \times 1$ query vector. Due to the properties of homomorphic encryption, this results in an $m \times 1$ encrypted vector that contains the desired document chunks from cluster $i$. This encrypted result is then sent back to the client.

\textbf{Client-Side Finalization.} The client decrypts the response from the server to obtain the plaintext documents of the target cluster. It then performs a final local re-ranking of these documents to identify and present the top-K most relevant results to the user.

\section{Evaluation}
To rigorously evaluate PIR-RAG, we compare it against our own implementations of two powerful baseline architectures, representing the state-of-the-art in graph-based and private-scoring paradigms.

\subsection{Experimental Setup}
\begin{itemize}
    \item \textbf{Datasets:} We use the MS MARCO document collection for quality evaluation and the SIFT1M dataset for scalability analysis.
    \item \textbf{Baselines:} We implemented two baseline systems for direct comparison:
        \begin{enumerate}
            \item \textbf{Graph-PIR:} An architecture inspired by systems like PACMANN \cite{pacmann2024}. It constructs a k-NN similarity graph over document embeddings. Retrieval is performed via a private, multi-step graph traversal using the same LHE-based PIR primitive to fetch neighbors at each hop.
            \item \textbf{Tiptoe-style Private Scoring:} An architecture based on the principles of Tiptoe \cite{tiptoe2023}. It uses clustering to partition data. The server receives an encrypted query and homomorphically computes similarity scores for all documents in the target cluster, returning only the encrypted scores. A key limitation is that the server knows which cluster the user is querying, which may leak information.
        \end{enumerate}
    \item \textbf{Metrics:} We measure Setup Time, Query Time, Communication, and Search Quality (NDCG, Precision, Recall).
\end{itemize}

\subsection{Scalability Analysis}
\begin{figure*}[ht!]
    \centering
    \begin{subfigure}[b]{0.48\textwidth}
        \centering
        \includegraphics[width=\linewidth]{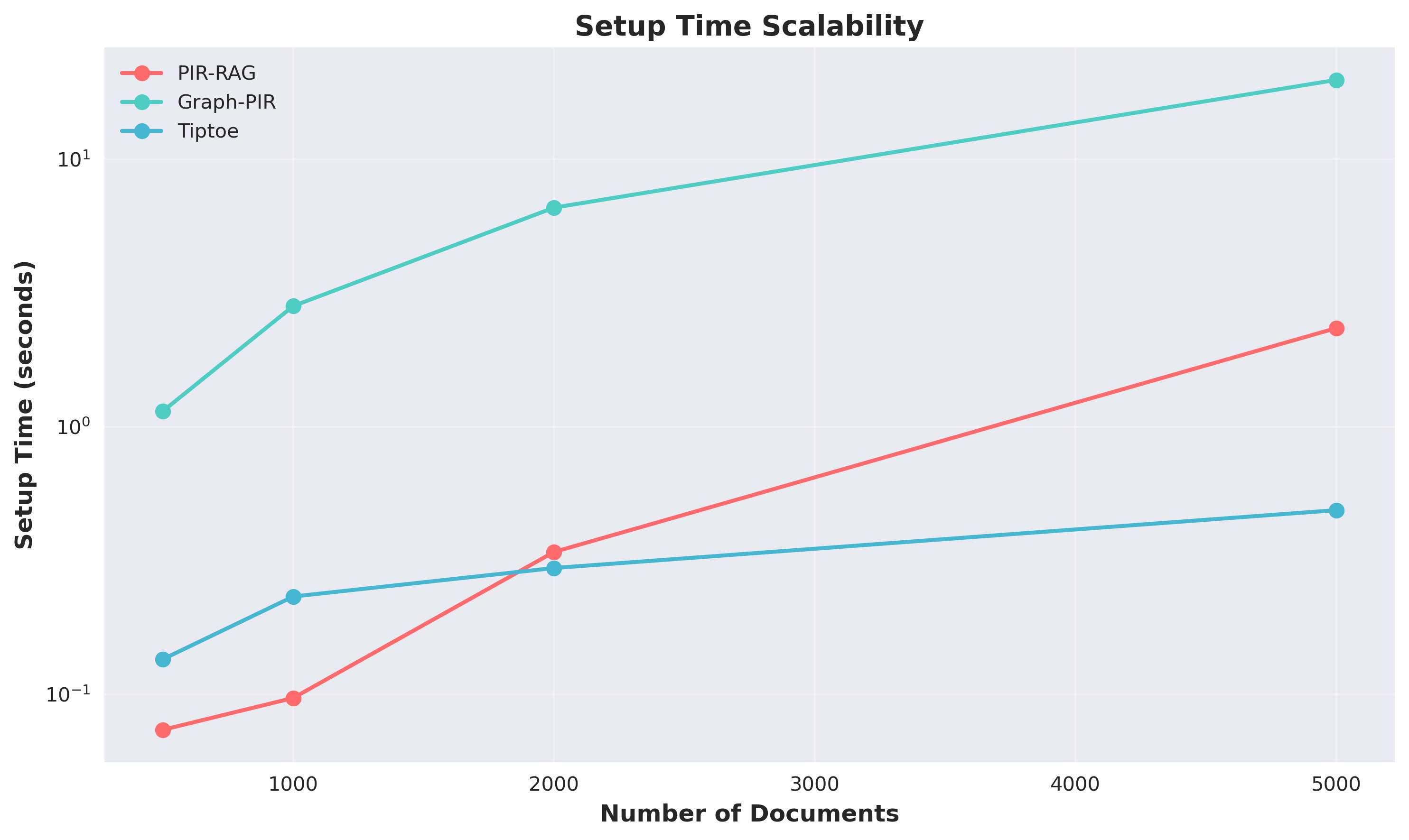}
        \caption{Setup Time vs. Document Size}
        \label{fig:setup_time_sub}
    \end{subfigure}
    \hfill
    \begin{subfigure}[b]{0.48\textwidth}
        \centering
        \includegraphics[width=\linewidth]{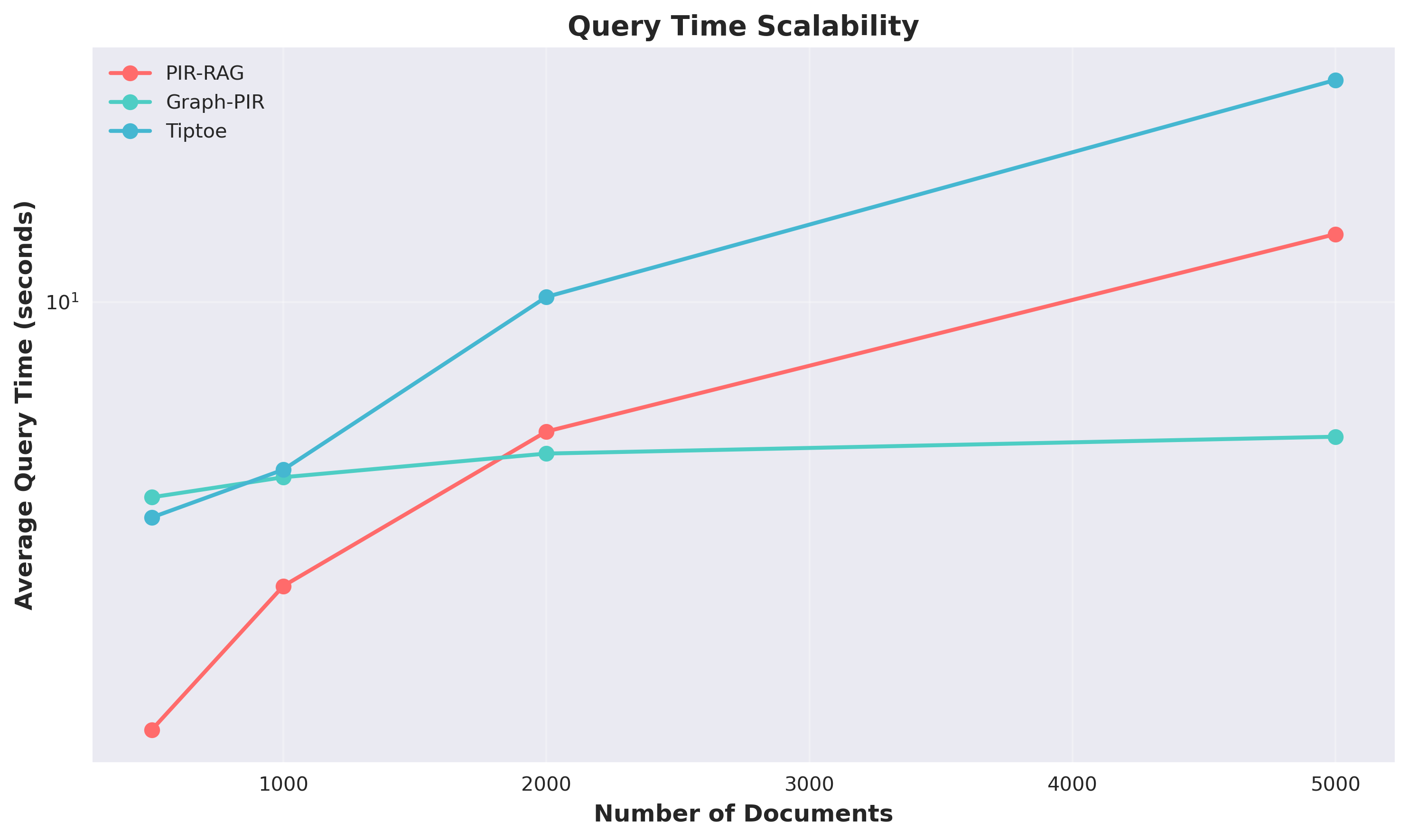}
        \caption{Query Time vs. Document Size}
        \label{fig:query_time_sub}
    \end{subfigure}
    
    \vspace{0.5cm} 
    
    \begin{subfigure}[b]{0.48\textwidth}
        \centering
        \includegraphics[width=\linewidth]{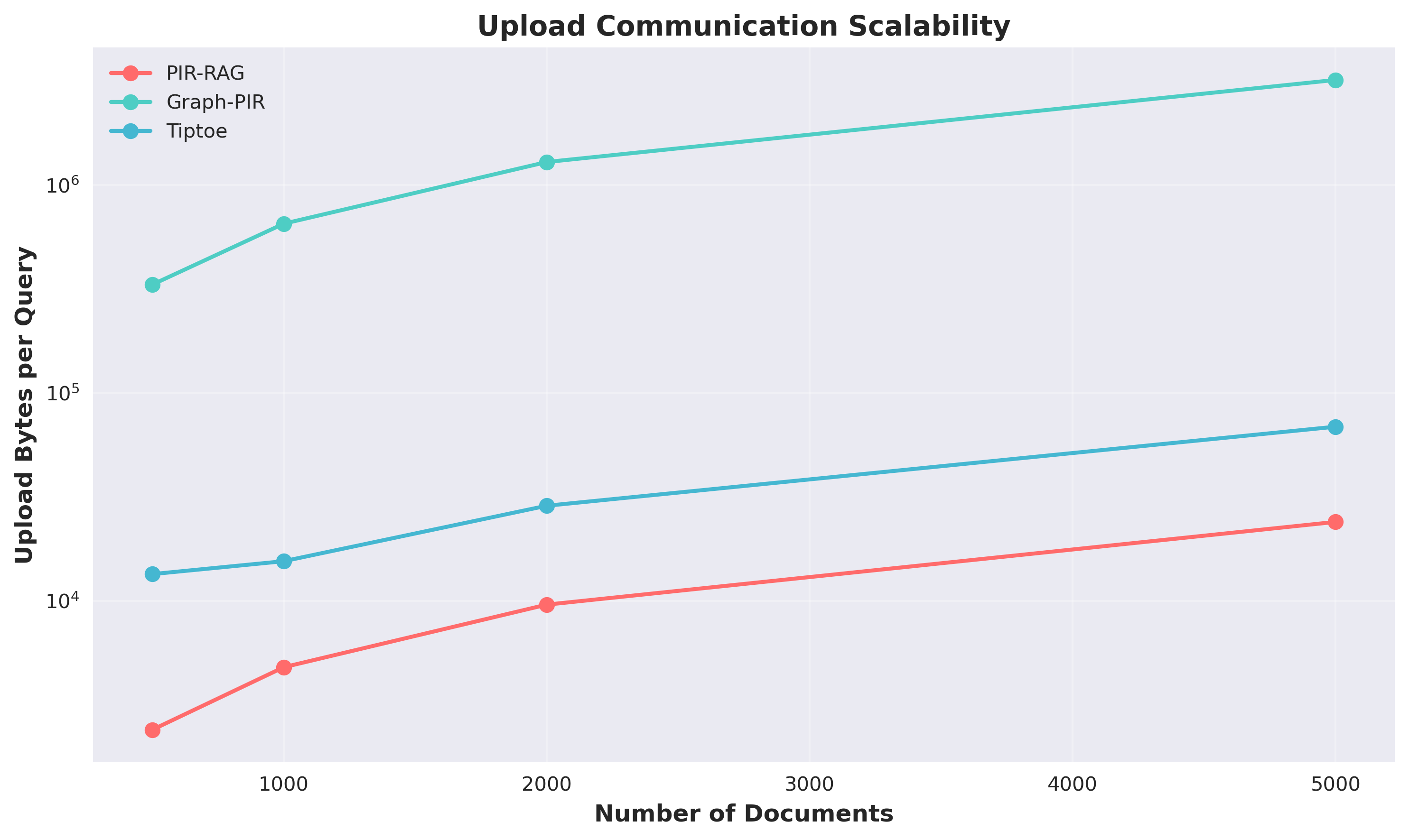}
        \caption{Upload Communication vs. Document Size}
        \label{fig:upload_sub}
    \end{subfigure}
    \hfill
    \begin{subfigure}[b]{0.48\textwidth}
        \centering
        \includegraphics[width=\linewidth]{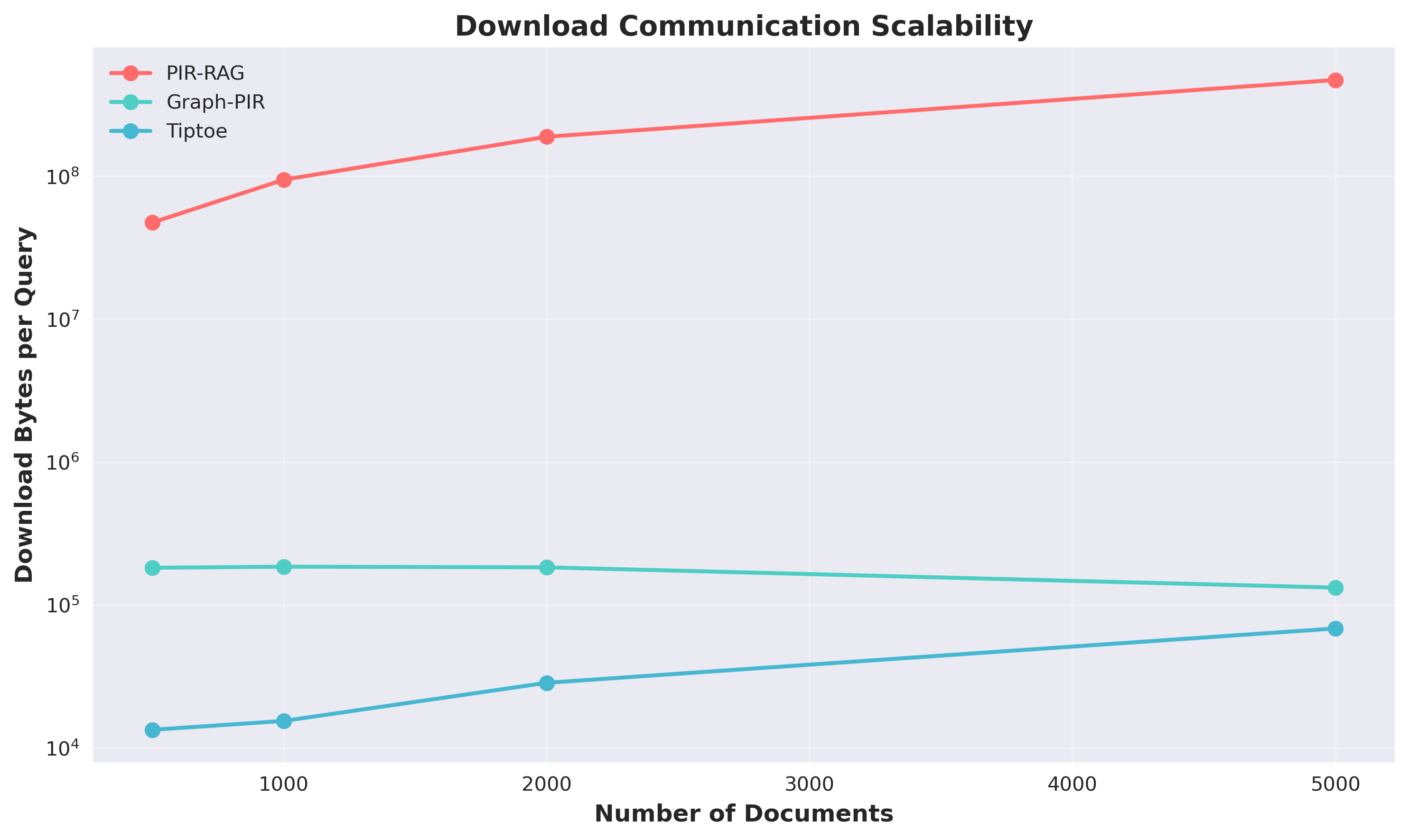}
        \caption{Download Communication vs. Document Size}
        \label{fig:download_sub}
    \end{subfigure}
    
    \caption{Scalability analysis of PIR-RAG, Tiptoe, and PACMANN across varying database sizes on the SIFT1M dataset. (a) One-time setup costs. (b) End-to-end query latency. (c) Uplink communication per query. (d) Downlink communication per query.}
    \label{fig:scalability_combined}
\end{figure*}

We evaluate how the performance of each system scales with the number of documents. The results are presented in Figure~\ref{fig:scalability_combined}.

The setup time analysis in Figure~\ref{fig:scalability_combined}(a) highlights the significant one-time cost of graph construction for Graph-PIR, which grows to nearly 20 seconds for 5,000 documents. In contrast, the cluster-based setups for PIR-RAG and the Tiptoe-style system are substantially faster.

Figure~\ref{fig:scalability_combined}(b) illustrates the query latency. Graph-PIR demonstrates excellent scalability, with its query time remaining relatively stable regardless of database size. Conversely, the query times for both PIR-RAG and the Tiptoe-style system show a clear linear growth, as their performance is tied to the increasing cluster size.

The communication costs are shown in Figures~\ref{fig:scalability_combined}(c) and \ref{fig:scalability_combined}(d). The downlink communication reveals the core architectural trade-offs: PIR-RAG incurs the highest cost by far (growing to nearly 475 MB), as it fetches the full plaintext content of an entire cluster. Graph-PIR and the Tiptoe-style system are far more conservative, transmitting only kilobytes of data. This stark difference is central to our analysis of the systems' fitness for the end-to-end RAG task.

Communication costs are shown in Figures~\ref{fig:scalability_combined}(c) and \ref{fig:scalability_combined}(d). PIR-RAG has the most efficient uplink communication (2.4 KB to 24 KB), as it only needs to transmit a single encrypted one-hot vector. The downlink communication reveals the core architectural trade-offs: PIR-RAG incurs the highest cost by far (growing from 47.6 MB to 474 MB), as it fetches the full plaintext content of an entire cluster. Graph-PIR and the Tiptoe-style system are far more conservative in their download, transmitting only a few hundred kilobytes, as they only return identifiers or scores. This stark difference in download cost is central to our analysis of the systems' fitness for the end-to-end RAG task.

\subsection{Search Quality and Performance Comparison}
\begin{figure*}[ht!]
    \centering
    
    \begin{subfigure}[b]{0.48\textwidth}
        \centering
        \includegraphics[width=\linewidth]{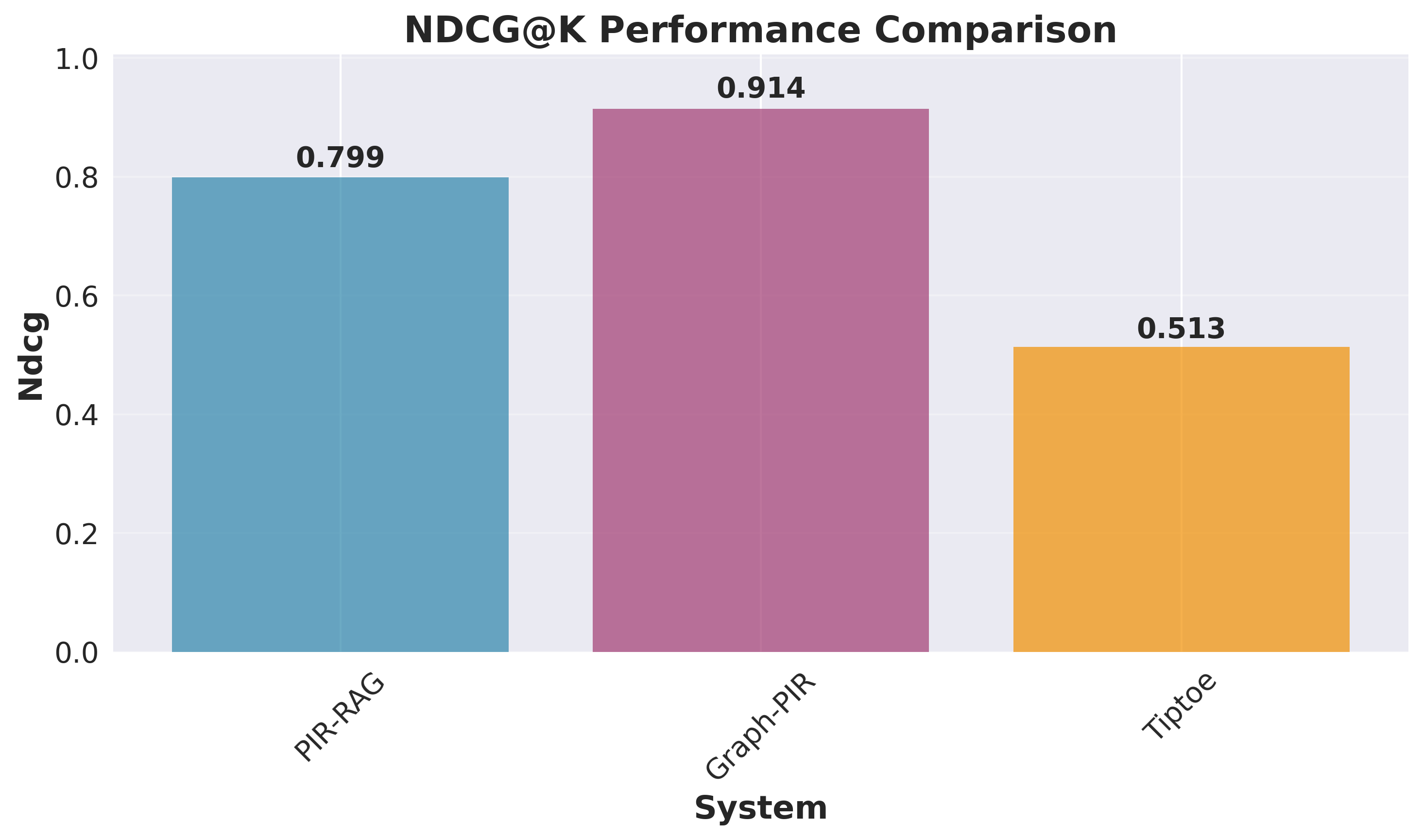}
        \caption{Comparison of NDCG}
        \label{fig:ndcg_sub}
    \end{subfigure}
    \hfill 
    \begin{subfigure}[b]{0.48\textwidth}
        \centering
        \includegraphics[width=\linewidth]{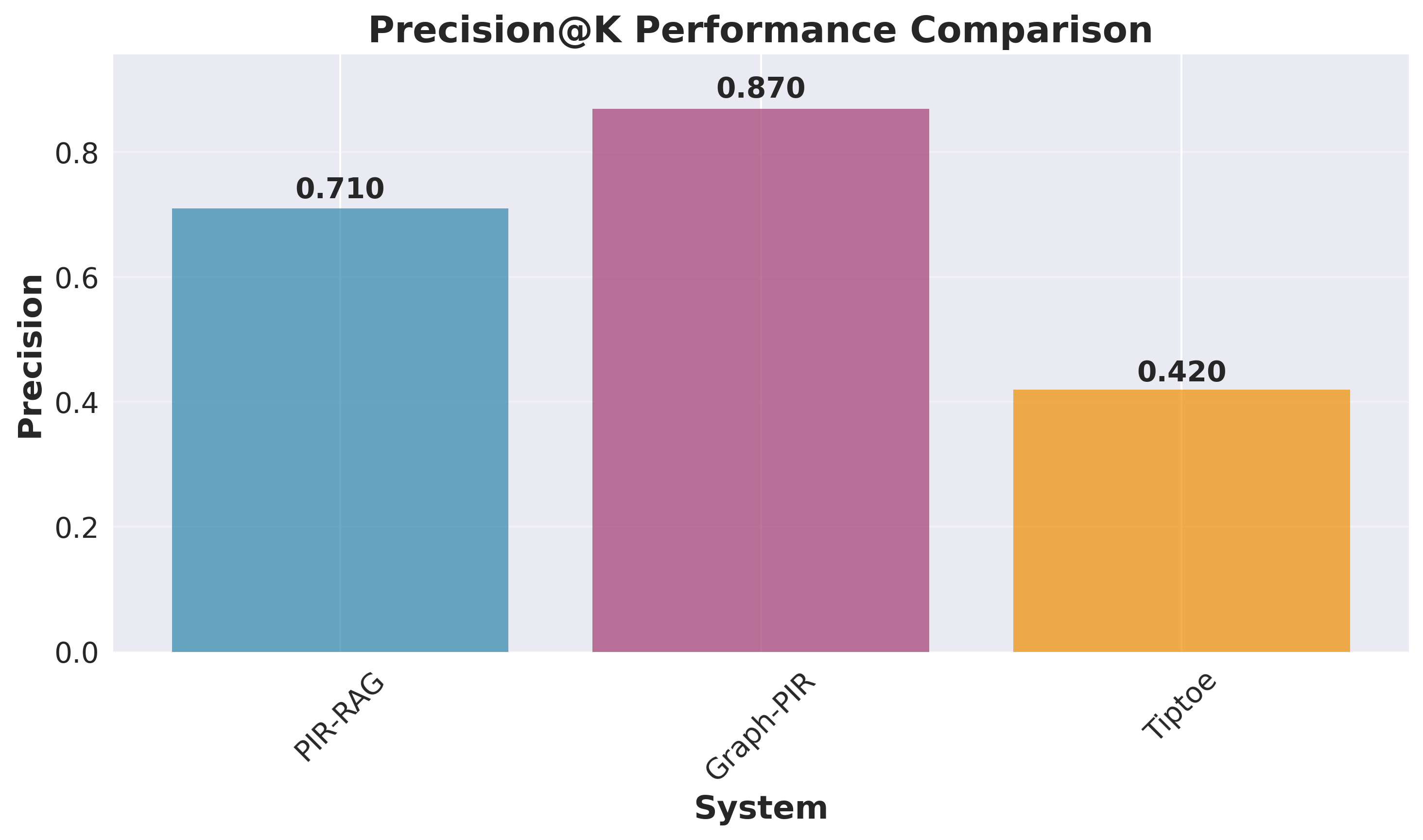}
        \caption{Comparison of Precision}
        \label{fig:precision_sub}
    \end{subfigure}
    
    \vspace{0.5cm} 
    
    \begin{subfigure}[b]{0.5\textwidth}
        \centering
        \includegraphics[width=\linewidth]{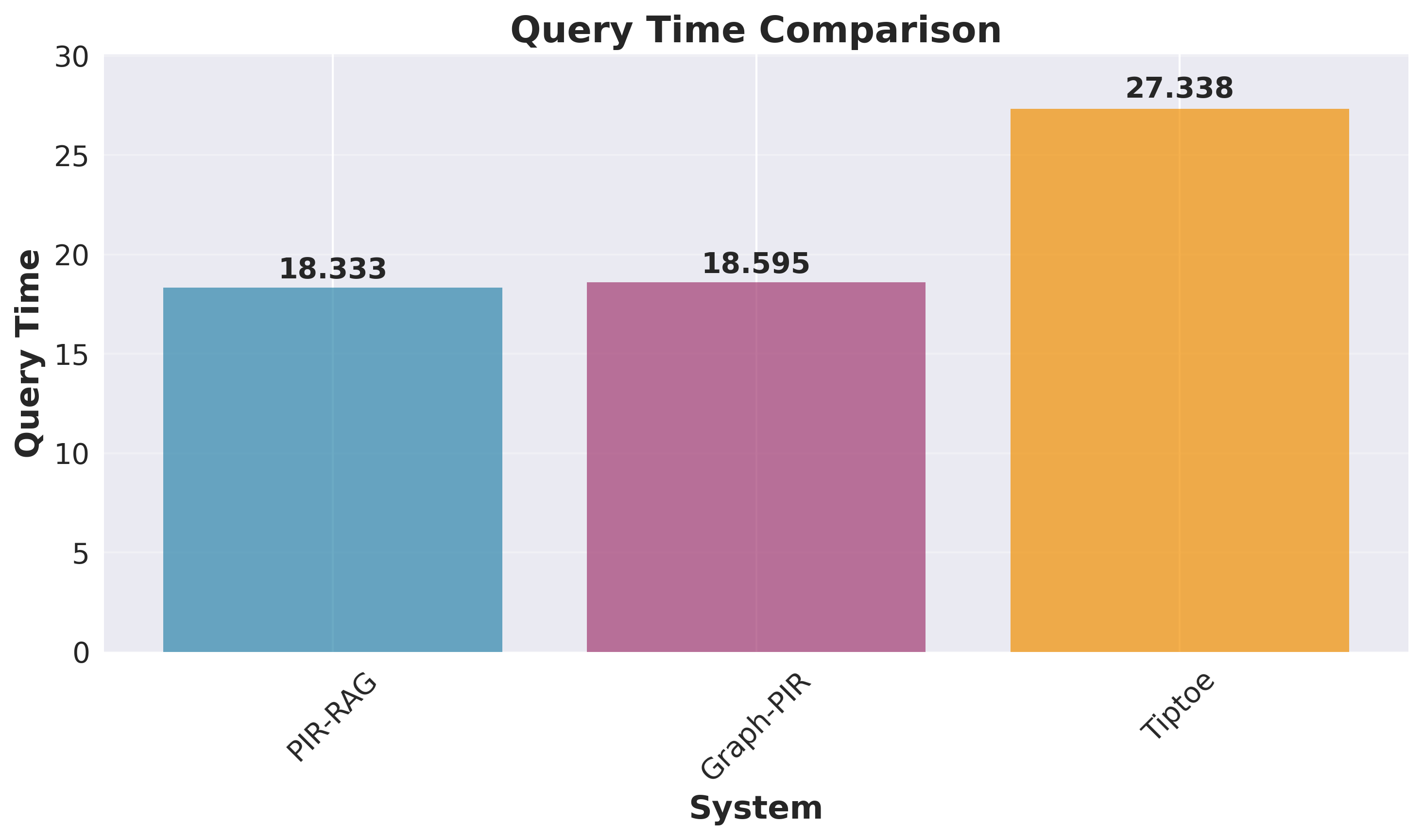}
        \caption{Comparison of Query Time}
        \label{fig:query_time_comp_sub}
    \end{subfigure}
    
    \caption{Performance and quality comparison on the MS MARCO dataset. (a) Normalized Discounted Cumulative Gain (NDCG) and (b) Precision results illustrate the search quality trade-offs. (c) End-to-end query time for the retrieval phase highlights the raw performance of each architecture.}
    \label{fig:quality_and_time}
\end{figure*}

In Figure~\ref{fig:quality_and_time}, we present a detailed comparison of search quality and query time on a fixed-size task of 5,000 documents from the MS MARCO dataset.

The search quality results, shown in Figures~\ref{fig:quality_and_time}(a) and \ref{fig:quality_and_time}(b), reveal a clear hierarchy. Graph-PIR achieves the highest search quality (0.901 NDCG@10 and 0.850 Precision@10), demonstrating the effectiveness of its fine-grained graph traversal. PIR-RAG provides strong, competitive quality (0.799 NDCG@10 and 0.710 Precision@10), significantly outperforming the Tiptoe-style architecture, which trails with the lowest quality (0.513 NDCG@10).

Figure~\ref{fig:quality_and_time}(c) compares the end-to-end query time for the retrieval phase. Here, Graph-PIR is the fastest at 12.99s, followed by PIR-RAG at 16.84s, and the Tiptoe-style system at 23.82s. However, this raw latency is deceptive for a full RAG workflow. For a true RAG application, both Graph-PIR and the Tiptoe-style system would require K additional, expensive PIR queries to privately fetch the document content for the LLM. In contrast, PIR-RAG's measured time already includes this content fetching step. This architectural efficiency in combining retrieval and fetching makes PIR-RAG's performance profile highly attractive for real-world RAG deployments, a concept we term "RAG-Ready Latency." When this full workflow is considered, PIR-RAG's single-shot "retrieve-and-fetch" design becomes significantly more efficient than the multi-step "retrieve-then-fetch" process required by the others.

\section{Conclusion}
We introduced and evaluated PIR-RAG, a privacy-preserving system for Retrieval-Augmented Generation. Our comprehensive comparison against two strong baseline architectures—graph-based PIR and Tiptoe-style private scoring—reveals a clear spectrum of design trade-offs. While graph-based methods offer superior raw retrieval accuracy, and private scoring offers fast but incomplete results, we conclude that PIR-RAG's "cluster-and-fetch" architecture is the most practically efficient solution for end-to-end private RAG tasks. By uniquely combining the retrieval and fetching phases into a single, batched PIR operation, it avoids the prohibitive costs of multiple, sequential private fetches required by other architectures. This work provides a clear, data-driven foundation for developers to select and build the next generation of powerful and privacy-respecting AI systems.

\bibliographystyle{plain}
\bibliography{ref}
\end{document}